\documentclass[11pt]{article}

\usepackage{epsfig}
\usepackage{subfig}
\usepackage{graphicx}
\usepackage{epstopdf}
\usepackage{amsfonts}
\usepackage{amssymb}
\usepackage{amsthm}
\usepackage{amsmath}
\usepackage{multirow}
\usepackage{color}
\usepackage{cite}
 \usepackage{makecell}
\def\beq{\begin{equation}}
\def\eeq{\end{equation}}
\def\bea{\begin{eqnarray}}
\def\eea{\end{eqnarray}}
\newcommand{\beqs}{\begin{subequations}}
\newcommand{\eeqs}{\end{subequations}}

\newcommand{\cref}[1]{Ref.~\cite{#1}}

\newcommand{\hh}{{\ensuremath{I{\kern-2.6pt h}}}}
\newcommand{\bhh}{{\ensuremath{\bar{I{\kern-2.6pt h}}}}}

\usepackage[colorlinks=true,allcolors=blue]{hyperref}

\begin{document}

\begin{titlepage}
	

\begin{center}
{\Large {\bf Monopoles, Strings and Gravitational Waves in Non-minimal Inflation}
}
\\[12mm]
Rinku Maji,$^{1}$
Qaisar Shafi$^{2}$~
\end{center}
\vspace*{0.50cm}
	\centerline{$^{1}$ \it
		Theoretical Physics Division, Physical Research Laboratory,}
		\centerline{\it  Navarangpura, Ahmedabad 380009, India}
	\vspace*{0.2cm}
	\centerline{$^{2}$ \it
		Bartol Research Institute, Department of Physics and 
		Astronomy,}
	\centerline{\it
		 University of Delaware, Newark, DE 19716, USA}
	\vspace*{1.20cm}
\begin{abstract}
We discuss how in $SO(10)$ grand unification an observable number density of topologically stable intermediate mass ( $\sim 10^{14}$ GeV) monopoles survive inflation driven by a Coleman-Weinberg potential and non-minimal coupling of the inflaton field to gravity.  The scalar spectral index $n_s$ is in excellent agreement with the current observations, and the tensor to scalar ratio $r\gtrsim 0.003$.
The model also predicts the presence of intermediate scale topologically stable cosmic strings, and their gravitational wave spectrum reflects the amount of cosmic inflation experienced by the associated symmetry breaking.
The discovery of these primordial monopoles and the stochastic gravitational wave background from the strings would provide important new insights regarding the symmetry breaking patterns in the early universe.
\end{abstract}

\end{titlepage}
\section{Introduction}
An early successful model of inflation based on a grand unified theory (GUT) employed a Coleman-Weinberg potential \cite{Coleman:1973jx} for a GUT singlet scalar inflaton field with minimal coupling to gravity \cite{Shafi:1983bd,Lazarides:1984pq}. It was soon followed by models which showed how topologically stable cosmic strings \cite{Kibble:1982ae,Shafi:1984tt} and intermediate scale monopoles \cite{Lazarides:1980cc,Rehman:2008qs,Senoguz:2015lba,Lazarides:2019xai} in GUTs such as $SO(10)$ may survive an inflationary epoch.  We do not include dimensionful parameters in the potential following the philosophy of Coleman-Weinberg potential as in \cite{Shafi:1983bd}. Coleman-Weinberg potential provides a nice framework for implementing a consistent inflationary scenario in grand unified theories. It allows us to implement non-minimal inflation starting with an unbroken GUT symmetry, and intermediate scale symmetries can be broken as the inflation rolls towards its minimum. 

Clearly, the experimental observations of monopoles and strings would have far reaching consequences for both particle physics and cosmology. For instance, if cosmic strings experience a significant amount of inflation without being inflated away, their subsequent emission of gravitational waves would  be modified relative to the standard scenario without any impact from inflation \cite{Cui:2019kkd,Chakrabortty:2020otp,Lazarides:2021uxv}.

Our main goal in this paper is to consider the impact on intermediate scale cosmic strings and monopoles from inflation driven by a GUT singlet scalar inflaton field with a Coleman-Weinberg potential and non-minimal coupling to gravity \cite{Spokoiny:1984bd,Bezrukov:2007ep,Okada:2010jf,Linde:2011nh, Martin:2013tda,Iso:2014gka, Galante:2014ifa,Campista:2017ovq,Tenkanen:2017jih,Oda:2017zul,Bostan:2018evz,Bostan:2019uvv,Bostan:2019fvk,Bostan:2020pnb,Kubo:2020fdd,Ghoshal:2022qxk,Okada:2022yvq}. To simplify things, we focus on an SO(10) model with two distinct symmetry breaking patterns that yield topologically stable monopoles and strings. The intermediate scales are determined by requiring unification of the standard model (SM) gauge couplings, which are then used to study the impact of partial inflation on the observability of monopoles and strings associated with these scales.

We highlight the parameter space that yields a monopole number density that varies between the MACRO bound \cite{Ambrosio:2002qq} and a few orders of magnitude lower. If the cosmic strings have comparable mass we show the corresponding gravitational wave spectra as a function of frequency. We also discuss the case where the string mass scale is lighter than the monopole scale.

This paper is organized as follows. In Sec.~\ref{sec:CW-infl} we discuss the predictions for an inflationary scenario driven by a real GUT singlet scalar with a Coleman-Weinberg potential and non-minimal coupling to gravity. Sec.~\ref{sec:pt-infl} discusses spontaneous symmetry breaking, formation and evolution of topological defects in this inflationary scenario, and their observational imprints including monopole flux and gravitational waves. In Sec.~\ref{sec:so10} we provide examples of two realistic $SO(10)$ models which predict topologically stable monopoles and cosmic strings, and our conclusions are summarized in Sec.~\ref{sec:summary}.
\section{Coleman-Weinberg Inflation with Non-minimal Coupling to Gravity}
\label{sec:CW-infl}
We focus here on inflation driven by a real GUT singlet scalar, $\phi$, with non-minimal coupling to gravity. The relevant part of the Lagrangian in Jordan frame is given by \cite{Callan:1970ze,Coleman:1973jx,Jackiw:1974cv,Freedman:1974ze}
\begin{align}\label{eq:CW-potential}
\frac{\mathcal{L}_J}{\sqrt{-g}} = \frac{1}{2}f(\phi) R - \frac{1}{2} g^{\mu\nu} \partial_\mu \phi \partial_\nu \phi  - V_J(\phi) \ .
\end{align}
Here, we use units with the reduced Planck mass $m_{\rm Pl} =1/\sqrt{8\pi G} = 2.44\times 10^{18}$ GeV ($G$ is the gravitational constant) set equal to one. 
\begin{align}
V_J(\phi) = A \phi^4 \left(\log \left(\frac{\phi}{M}\right) - \frac{1}{4}\right) + \frac{AM^4}{4} \ ,
\end{align}
and
\begin{align}
f(\phi) = 1 + \xi (\phi^2 - M^2) ,
\end{align}
such that $f(\phi)\to 1$ as $\phi$ rolls to its minimum $M$ after inflation.

We apply Weyl rescaling $g^{\mu\nu}\to g_E^{\mu\nu} = f(\phi)g^{\mu\nu}$ to switch to the Einstein frame, and the Lagrangian in the Einstein frame reads \cite{Fujii:2003pa}
\begin{align}
\frac{\mathcal{L}_J}{\sqrt{-g_E}} = \frac{1}{2} R_E - \frac{1}{2}g_E^{\mu\nu}\left(\frac{1}{f(\phi)}+\frac{3}{2}\frac{f'(\phi)^2}{f(\phi)^2}\right) \partial_\mu \phi \partial_\nu \phi  - V(\phi)
\end{align}
where 
\begin{align}
V(\phi) = \frac{V_J(\phi)}{f(\phi)^2}
\end{align}
is the Einstein frame potential. We redefine the field as
\begin{align}\label{eq:sigmap}
\left(\frac{d\sigma}{d\phi}\right)^2 = \frac{1}{f(\phi)}+\frac{3}{2}\frac{f'(\phi)^2}{f(\phi)^2} = \frac{1+\xi(\phi^2 - M^2) + 6\xi^2\phi^2}{(1  + \xi (\phi^2 - M^2))^2} ,
\end{align}
and obtain a Lagrangian for a scalar $\sigma$ minimally coupled to gravity with a canonical kinetic term. We can define the slow-roll parameters as follows (see Ref.~\cite{Lyth:2009zz} and the references therein for a review):
\begin{align}
\epsilon = \frac{1}{2}\left( \frac{V_\sigma}{V}\right)^2, \quad \eta = \frac{V_{\sigma\sigma}}{V}, \quad \zeta = \frac{V_\sigma V_{\sigma\sigma\sigma }}{V^2}
\end{align}
where the subscripts denote derivatives with respect to $\sigma$.

The inflationary predictions scalar spectral index ($n_s$), tensor to scalar ratio ($r$), and  the running of the spectral index ($\alpha = dn_s/d\ln k$) are expressed as
\begin{align}
n_s = 1-6\epsilon +2\eta , \quad r = 16\epsilon , \quad \alpha = 16\epsilon\eta - 24\epsilon^2 -2 \zeta \ .
\end{align}
The curvature perturbation is given by,
\begin{align}
\Delta_R^2 = \left.\frac{V}{24\pi^2\epsilon}\right|_{k=k_*}
\end{align}
where $k_*$ is the pivot scale. The experimental values at 68\% confidence level at the pivot scale $k_* = 0.05~\mathrm{Mpc}^{-1}$ \cite{Planck:2018jri,BICEP:2021xfz} are given in Table~\ref{tab:exp-param-infl}.
\begin{figure}[h!]
\centering
\includegraphics[width=0.7\linewidth]{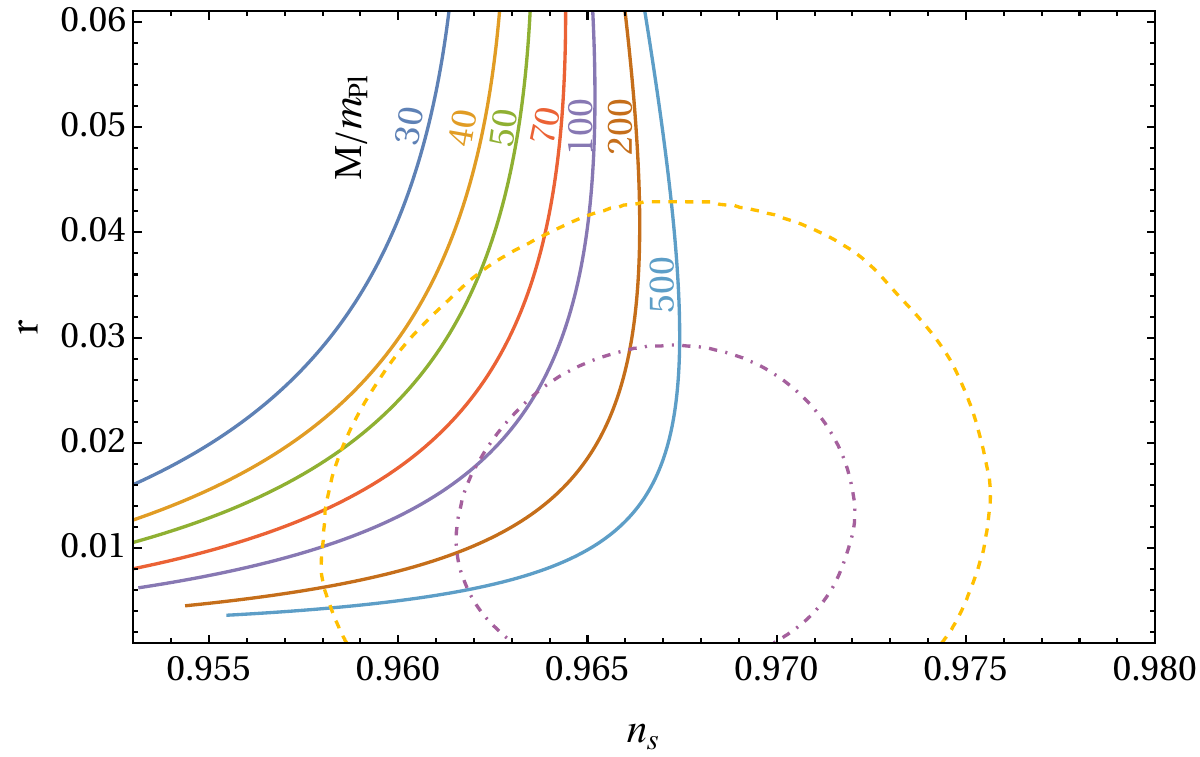}
\caption{Predictions for the scalar spectral index ($n_s$) and tensor to scalar ratio ($r$) in an inflationary scenario driven by the Coleman-Weinberg potential of a real GUT-singlet scalar with non-minimal coupling to gravity for different choices of $M$. We have set $T_r = 10^7$ GeV and $\omega_r=0$. The dot-dashed and dashed contours are the 68\% and 95\% confidence level contours of Planck TT, TE, EE $+$ lowE $+$ lensing $+$ BK18 $+$ BAO \cite{BICEP:2021xfz,Tristram:2021tvh}. Values of $r$ can range between 0.04 and 0.004.}
\label{fig:ns_r}
\end{figure}
\begin{table}
\begin{center}
\begin{tabular} { c  c}
\hline
 Parameter &  68\% limits\\
\hline
{\boldmath$\Delta_R^2$} & $(2.1\pm 0.1)\times 10^{-9}$\\

{\boldmath$n_s$} & $0.9669\pm 0.0037 $\\

{\boldmath$r$} & $0.0163^{+0.0061}_{-0.013} $\\
\hline
\end{tabular}
\end{center}
\caption{Experimental values \cite{BICEP:2021xfz} of inflation parameters at 68\% confidence level at the pivot scale $k_* = 0.05~\mathrm{Mpc}^{-1}$.}
\label{tab:exp-param-infl}
\end{table}

 The number of $e$-folds is given by,
\begin{align}\label{eq:N_1}
N_* = \frac{1}{\sqrt{2}}{\rm sgn(V')}\int_{\phi_e}^{\phi_*}d\phi \frac{\sigma'}{\sqrt{\epsilon(\phi)}}
\end{align}
where the prime denotes the derivative with respect to $\phi$ and the end of inflation occurs at $\phi_e$ with max$(|\eta|,\epsilon)\simeq 1$.

\begin{figure}[h!]
\centering
\includegraphics[width=0.7\linewidth]{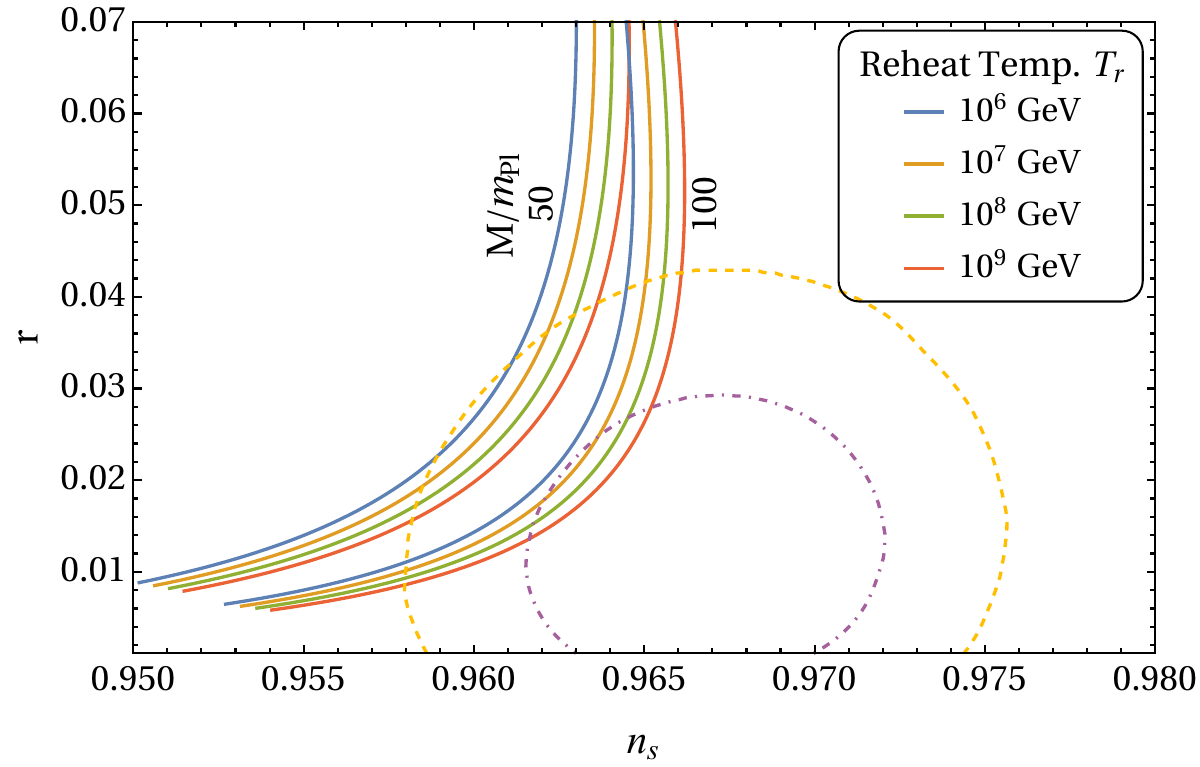}
\caption{Variation in the predictions for the scalar spectral index ($n_s$) and tensor to scalar ratio ($r$) with the reheat temperature ($T_r$) for two choices $M = 50 m_{\rm Pl}$ and $ M=100 m_{\rm Pl}$. The dot-dashed and dashed contours are the 68\% and 95\% confidence level contours of Planck TT, TE, EE $+$ lowE $+$ lensing $+$ BK18 $+$ BAO.}
\label{fig:ns_r_Tr}
\end{figure}
\begin{figure}[h!]
\centering
\includegraphics[width=0.7\linewidth]{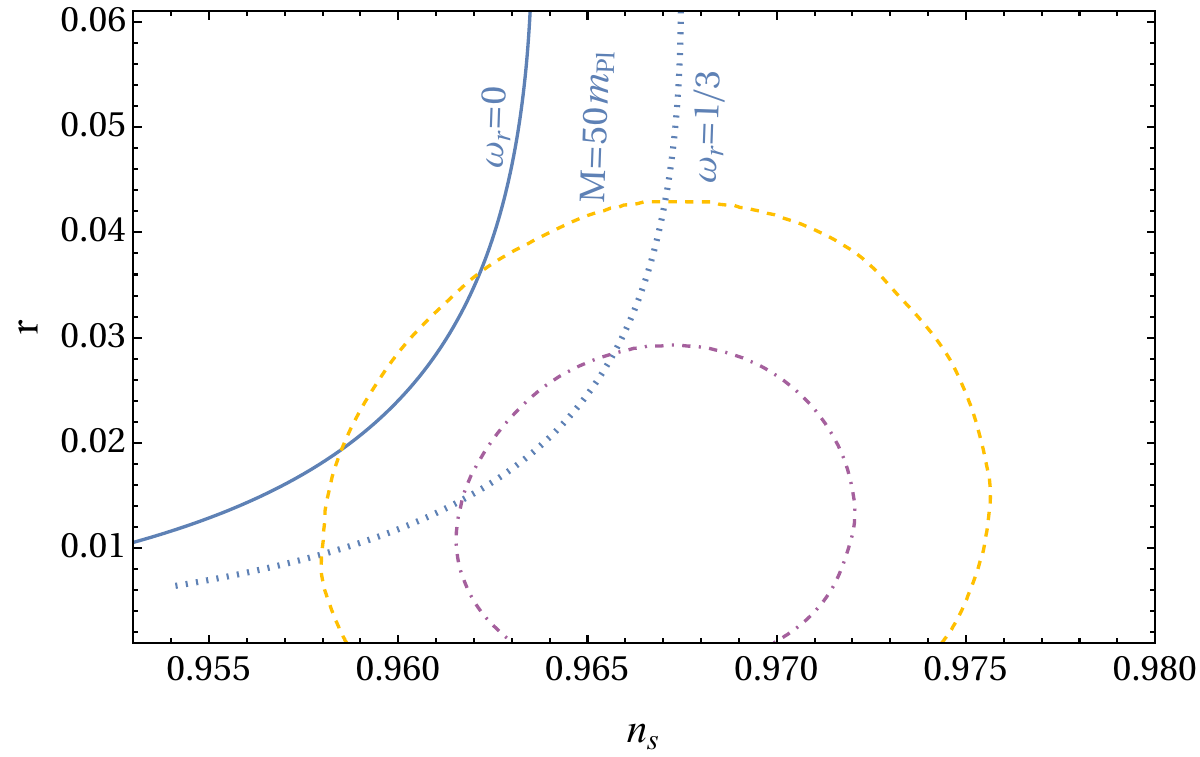}
\caption{Predictions for the scalar spectral index ($n_s$) and tensor to scalar ratio ($r$) with $M = 50 m_{\rm Pl}$ for the effective state parameter $w_r=0$ (solid line) and $w_r=1/3$ (dotted line). We have set $T_r = 10^7$ GeV. The dot-dashed and dashed contours are the 68\% and 95\% confidence level contours of Planck TT, TE, EE $+$ lowE $+$ lensing $+$ BK18 $+$ BAO.}
\label{fig:ns_r_wr}
\end{figure}

\begin{figure}[h!]
\begin{center}
\subfloat[]{\includegraphics[width=0.47\textwidth]{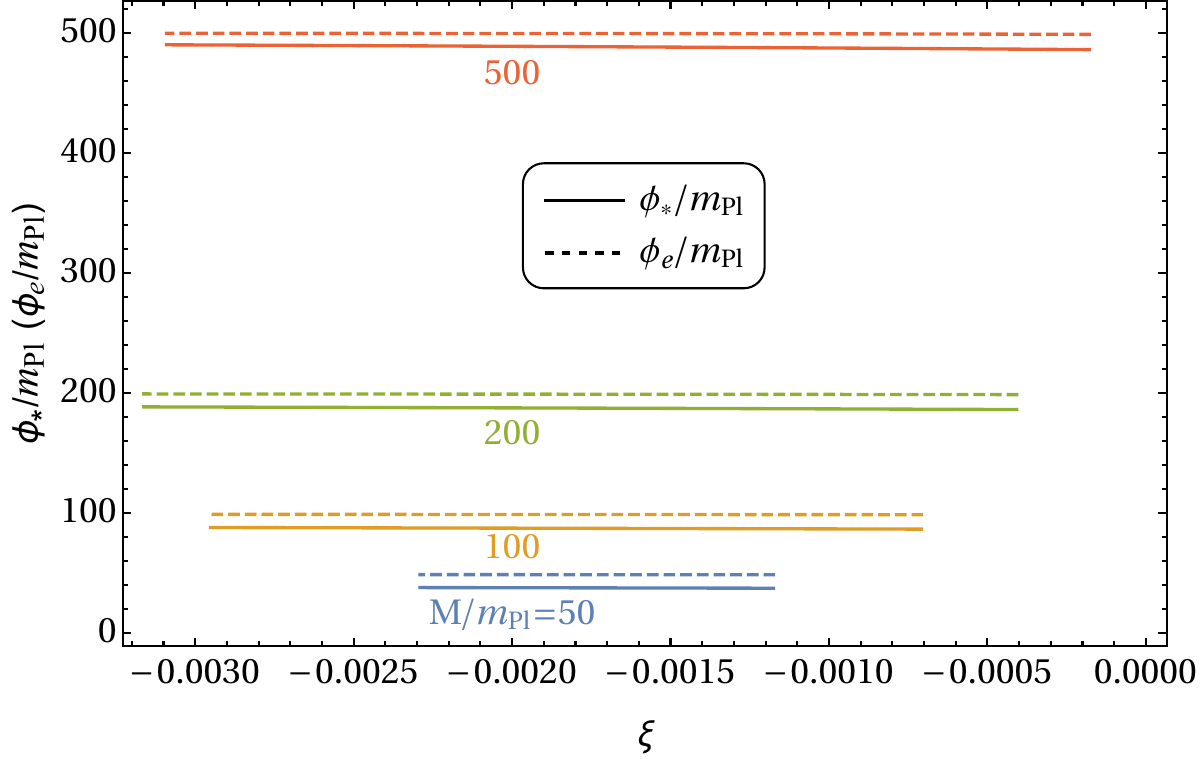}}
\subfloat[]{\includegraphics[width=0.5\textwidth]{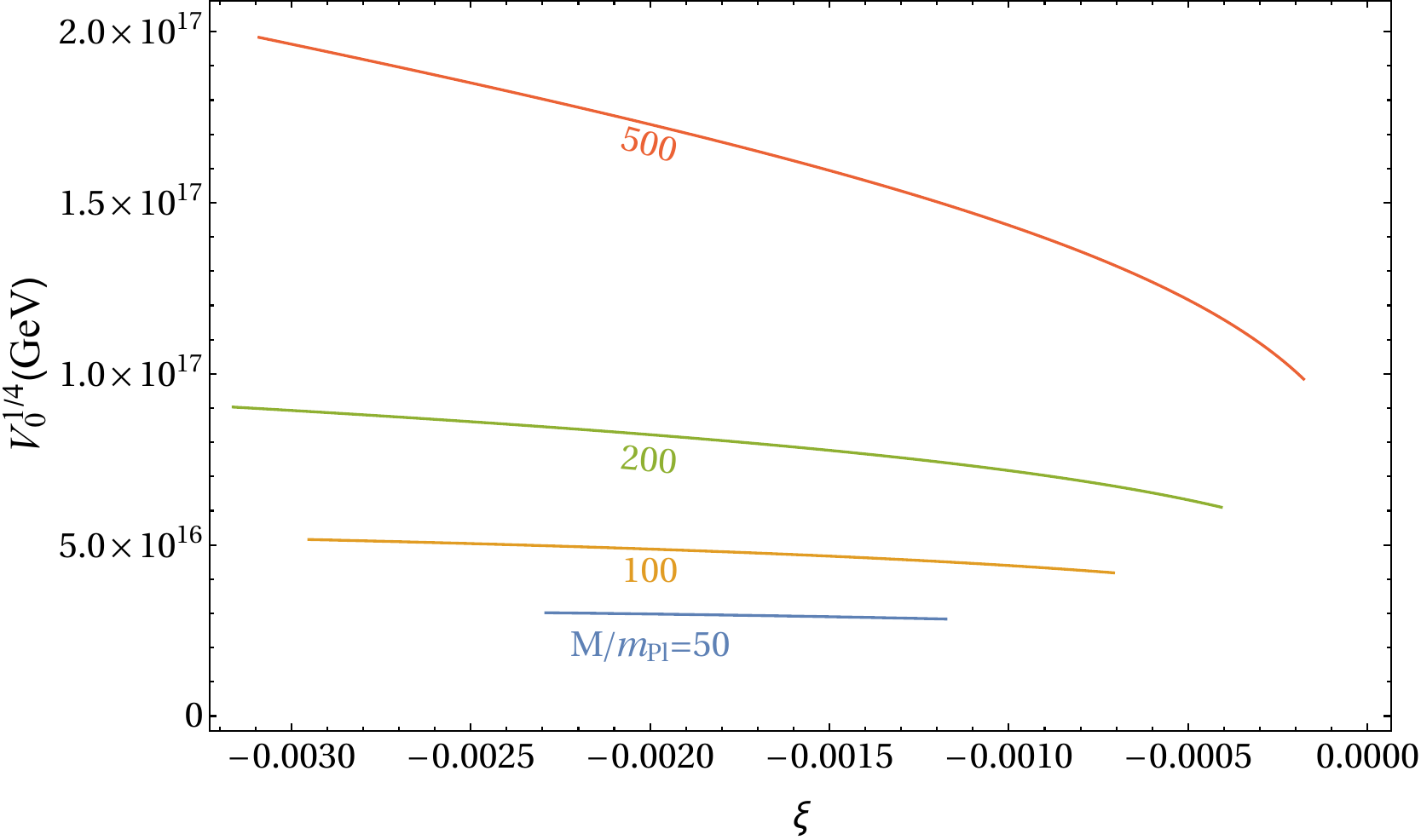}} \\
\subfloat[]{\includegraphics[width=0.47\textwidth]{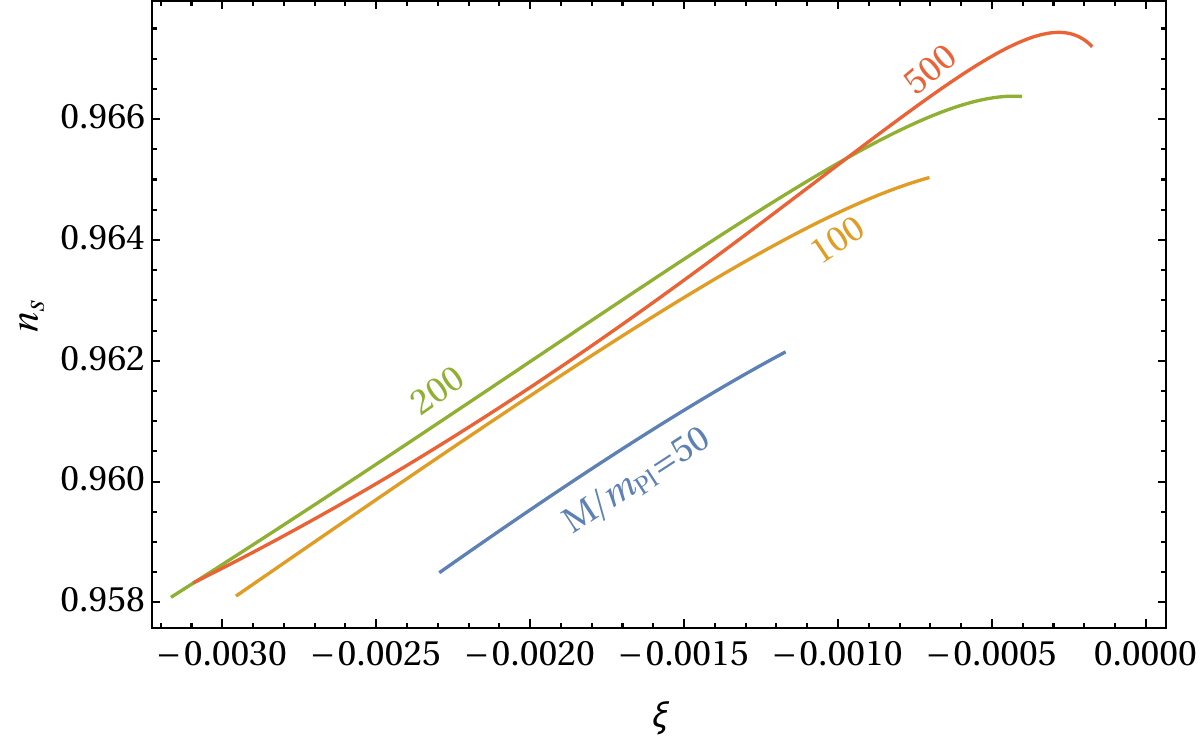}}
\subfloat[]{\includegraphics[width=0.47\textwidth]{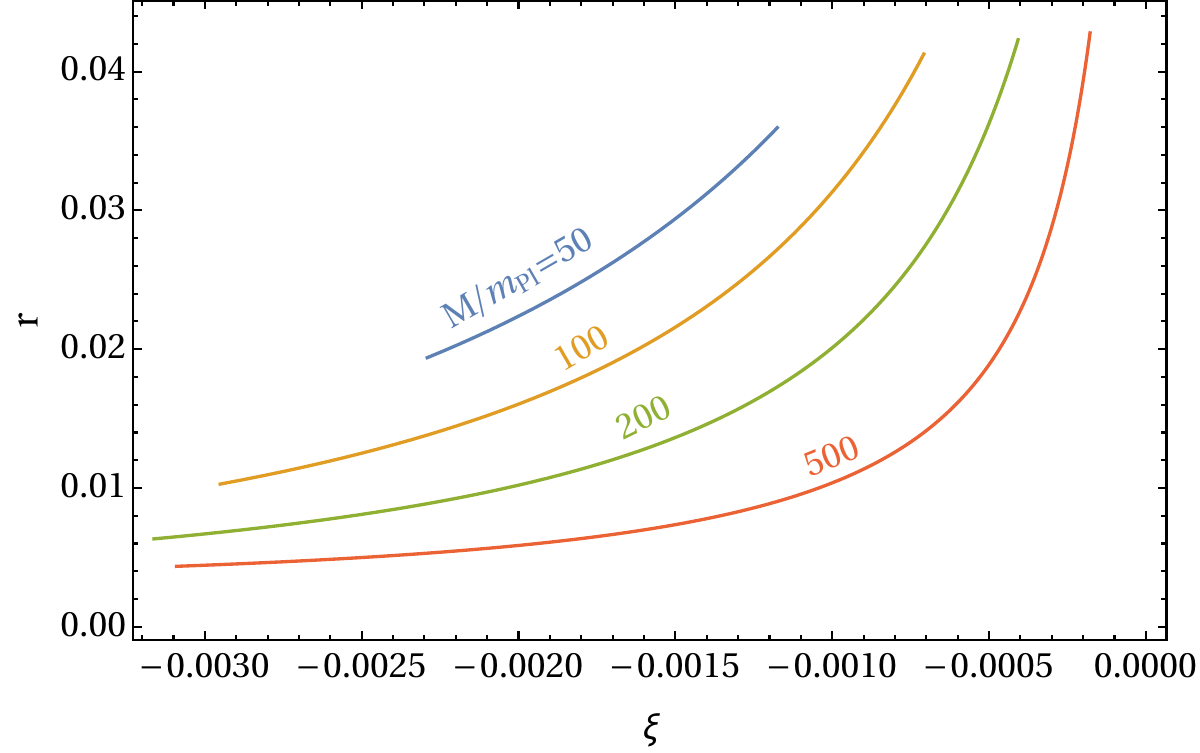}}
\end{center} 
\caption{Variation of different parameters in the inflaton potential as a function of $\xi$ for different values of $M$ for successful inflation compatible with Planck TT, TE, EE $+$ lowE $+$ lensing $+$ BK18 $+$ BAO. We have set $T_r = 10^7$ GeV and $\omega_r=0$.}
\label{fig:param-xi}
\end{figure}

\begin{table}[h!]
\begin{center}
\begin{tabular}{| c | c | c | c | c | c | c | c | c | c | c |}
\hline
$\xi$ & $\frac{M}{m_{\rm Pl}}$ & \thead{$\log_{10}\frac{V_0^{1/4}}{\mathrm{GeV}}$} & -$\log_{10}A$ & $\phi_*/m_{\rm Pl}$ & $\phi_e/m_{\rm Pl}$ & $n_s$ & $r$ & -$10^4\alpha$ & $\frac{10^{-13}H_*}{\mathrm{GeV}}$ & $N_*$ \\
\hline
\multirow{5}{*}{\rotatebox[origin=c]{90}{-0.001}}& 50 & 16.45 & 13.96 & 37.42 & 48.71 & 0.9626 & 0.040 & 6.51 & 4.99 & 50.6 \\
& 70 & 16.54 & 14.17 & 57.09 & 68.73 & 0.9637 & 0.037 & 6.46 & 4.78 & 50.5 \\
& 100 & 16.64 & 14.37 & 86.90 & 98.77 & 0.9644 & 0.031 & 6.40 & 4.39 & 50.4 \\
& 200 & 16.86 & 14.73 & 186.90 & 198.90 & 0.9653 & 0.020 & 6.31 & 3.52 & 50.2 \\
& 500 & 17.16 & 15.12 & 487.40 & 499.20 & 0.9652 & 0.010 & 6.40 & 2.53 & 49.9 \\
\hline
\multirow{5}{*}{\rotatebox[origin=c]{90}{-0.002}}& 50 & 16.47 & 13.84 & 37.90 & 48.79 & 0.9595 & 0.022 & 6.23 & 3.71 & 50.3 \\
& 70 & 16.58 & 14.02 & 57.64 & 68.84 & 0.9607 & 0.020 & 6.21 & 3.47 & 50.2 \\
& 100 & 16.69 & 14.19 & 87.52 & 98.92 & 0.9614 & 0.016 & 6.21 & 3.14 & 50.1 \\
& 200 & 16.92 & 14.49 & 187.70 & 199.10 & 0.9620 & 0.010 & 6.28 & 2.51 & 49.9 \\
& 500 & 17.24 & 14.79 & 488.80 & 499.50 & 0.9616 & 0.006 & 6.62 & 1.90 & 49.7 \\
\hline
\multirow{5}{*}{\rotatebox[origin=c]{90}{-0.003}}& 50 & 16.49 & 13.79 & 38.35 & 48.86 & 0.9559 & 0.014 & 6.06 & 2.95 & 50.1 \\
& 70 & 16.60 & 13.94 & 58.13 & 68.93 & 0.9571 & 0.012 & 6.08 & 2.75 & 50.1 \\
& 100 & 16.71 & 14.09 & 88.07 & 99.04 & 0.9579 & 0.010 & 6.12 & 2.49 & 50.0 \\
& 200 & 16.95 & 14.35 & 188.40 & 199.30 & 0.9586 & 0.007 & 6.30 & 2.03 & 49.8 \\
& 500 & 17.29 & 14.57 & 490.00 & 499.60 & 0.9586 & 0.004 & 6.77 & 1.65 & 49.7 \\
\hline
\end{tabular}
\caption{Typical parameter values for inflation with the Coleman-Weinberg potential of a real singlet scalar with non-minimal coupling to gravity. We have set $T_r = 10^7$ GeV and $\omega_r=0$.}
\label{tab:infl-para}
\end{center}
\end{table}
The number of $e$-foldings corresponding to the pivot scale $k_*=0.05 \ \mathrm{Mpc^{-1}}$ obtained by assuming a standard thermal history after inflation is given by \cite{Liddle:2003as},
\begin{equation}\label{eq:N_2}
N_* \simeq 61.5 + \frac{1}{2} \mathrm{ln} \frac{\rho_*}{m_{\rm Pl}^4}-\frac{1}{3(1+\omega_r)} \mathrm{ln} \frac{\rho_e}{m_{\rm Pl}^4} + \left(\frac{1}{3(1+\omega_r)} - \frac{1}{4} \right)\mathrm{ln} \frac{\rho_r}{m_{\rm Pl}^4} \ .
\end{equation}
Here $\rho_* = V(\phi_*)$, $\rho_e = V(\phi_e)$, and $\rho_r = (\pi^2/30) g_*T_r^4$ are the energy densities at the pivot scale, end of inflation, and at reheating temperature, respectively, $\omega_r$ is the effective equation-of-state parameter from the end of inflation until reheating, and $g_*$ is the effective number of relativistic degrees of freedom which is $106.75$ after reheating for the SM particles. The values of $N_*$ from Eqs.~\eqref{eq:N_1} and \eqref{eq:N_2} should agree, and the predicted values of $\Delta_R^2$, $n_s$, $r$ and $\alpha$ must satisfy the experimental data for successful inflation.

Fig.~\ref{fig:ns_r} shows the predictions for $n_s$ and $r$ for various choices of the parameters $\xi$ and $M$, and we have taken $\omega_r = 0$ and $T_r = 10^7$ GeV. The predictions for $n_s$ and $r$ change as we vary the reheat temperature $T_r$ and the effective state parameter $\omega_r$. In Fig.~\ref{fig:ns_r_Tr} we show this for various values of $T_r = \lbrace 10^6, 10^7, 10^8, 10^9\rbrace$ GeV, $\omega_r = 0$ and $M = \lbrace 50, 100  \rbrace m_{\rm Pl}$. In Fig.~\ref{fig:ns_r_wr} we have compared the $n_S$-$r$ predictions for $\omega_r=0$ (mid-$N$) and $\omega_r=1/3$ (high-$N$) cases. Fig.~\ref{fig:param-xi} displays the variation of the inflaton field $\phi_*$ ($\phi_e$) at the pivot scale (end of inflation when max$[\epsilon, \eta] = 1$), $V_0=AM^4/4$, $n_s$, and $r$ as a function of $\xi$ for different choices of $M$ for successful inflation compatible with Planck TT, TE, EE $+$ lowE $+$ lensing $+$ BK18 $+$ BAO \cite{BICEP:2021xfz}.

 Table~\ref{tab:infl-para} lists the prediction of different parameters for successful inflation. Here we have taken $\omega_r = 0$ and $T_r = 10^7$ GeV. Whereas, the CW potential with minimal coupling to gravity is in tension with the latest data, we have identified a perfectly compatible window below the vacuum expectation value (VEV) for $M >> m_{\rm Pl}$ and $|\xi|$ around $10^{-3}$. This is close to the results obtained for inflation with a quadratic potential and non-minimal coupling to gravity, see \cite{Tenkanen:2017jih}.
\section{Symmetry Breaking, Topological Defects and Non-minimal Inflation}
\label{sec:pt-infl}
The part of the potential that dictates the interaction of the inflaton $\phi$ with some gauge symmetry breaking scalar $\chi_D$ is given by
\begin{align}\label{eq:inter-pot}
V_J(\phi,\chi_D) = - \frac{1}{2}\beta_D^2\phi^2\chi_D^2 + \frac{\lambda_D}{4}\chi_D^4, 
\end{align}
where, for simplicity, $\chi_D$ is a canonically normalized real scalar field in $D$-dimensional representation of the gauge group. Bare mass parameters are not allowed in our scenario by invoking scale invariance for both the gauge singlet and gauge-charged fields. By excluding such terms we are able to arrange the number of $e$-foldings experienced by a specific symmetry breaking.
The potential in Eq.~\eqref{eq:inter-pot} induces a vacuum expectation value (VEV) for $\chi_D$ given by
\begin{align}\label{eq:vev-final}
\left<\chi_D\right> = (\beta_D/\sqrt{\lambda_D}) M ,
\end{align}
when the inflaton reaches its minimum at $M$.
The coefficient $A=\beta_D^4\, D/16\pi^2$ \cite{Lazarides:2019xai} in Eq.~(\eqref{eq:CW-potential}) is dominated by the coupling of the GUT-breaking scalar and the unification scale is $M_U = \sqrt{\frac{8\pi}{\lambda_D}}\left(\frac{AM^4}{D}\right)^{1/4}$.


During inflation, the scalar $\chi_D$ has an effective mass squared given by
\begin{align}\label{eq:effec-mass}
m_{\rm eff}^2 = 2 (\beta_D^2\phi^2 - \sigma_{\chi_D} T_H^2) ,
\end{align}
where $T_H=H/2\pi$ is the Hawking temperature and we take $\sigma_{\chi_D} \sim 1$. 
The Ginzburg criterion \cite{ginzburg} which governs the onset of the phase transition is given by,
\begin{align}\label{eq:Ginz}
\xi_G^3 \Delta V > T_H ,
\end{align}
  where $\xi_G \sim \mathrm{min} [ H^{-1}, m_{\rm eff}^{-1} ]$ is the correlation length and $\Delta V=m_{\rm eff}^4/(16\lambda_D)$ is the difference between the potential at  $\chi_D = 0$ and $\left< \chi_D \right>$. 
From Eqs.~\eqref{eq:vev-final}, \eqref{eq:effec-mass} and \eqref{eq:Ginz} we can write the symmetry breaking scale as
\begin{align}\label{eq:breaking-scale}
M_I\sim \left<\chi_D\right> = \begin{cases}
 \sqrt{\left(128 \lambda_D^2 + \sigma_{\chi_D}\right)} \frac{M}{\sqrt{\lambda_D}\phi_I} \frac{H_I}{2\pi} , & \mathrm{for} \ \ m_{\rm eff}^{-1} \leq H^{-1} , \\
 \sqrt{\left(4\pi\sqrt{2\pi \lambda_D} + \sigma_{\chi_D}\right)} \frac{M}{\sqrt{\lambda_D}\phi_I} \frac{H_I}{2\pi} , & \mathrm{for} \ \ m_{\rm eff}^{-1} \geq H^{-1} ,
 \end{cases}
\end{align}
with the subscript $I$ denotes the appropriate intermediate scale values. 

The number density of monopoles at formation is of order $\xi_G^{-3}$ which will be diluted by a factor of $\exp(-3N_I)$ during inflation, and by another factor of $(\tau/t_r)^2$ during the inflaton oscillation from the end of inflation at time $\tau$ till the reheat time $t_r\simeq\Gamma_\phi^{-1}$ ($\Gamma_\phi$ is the decay width of $\phi$). The comoving number density the  so-called yield of the monopoles after the reheating is given by
\begin{align}\label{eq:Y_M}
Y_M \simeq \frac{\xi_G^{-3} \exp(-3N_I)\left(\frac{\tau}{t_r}
\right)^2}{\frac{2\pi^2}{45}g_{*}T_r^3} ,
\end{align}
where $N_I$ is obtained from Eq.~\eqref{eq:N_1} with $\phi_*$ replaced by $\phi_I$. From the Klein-Gordon (KG) equation for the inflaton field $\sigma$ ($\sigma$ is defined in Eq.~\eqref{eq:sigmap})
\begin{align}
3H\dot{\sigma}+\frac{\partial V}{\partial\sigma}\simeq 0 \ ,
\end{align}
we estimate the cosmic time at the end of inflation as
\begin{align}\label{eq:tend}
\tau \simeq\int_{\phi_e}^{\phi_*} \frac{3H(\phi)}{V'}\left(\frac{d\sigma}{d\phi}\right)^2 d\phi \ .
\end{align}
\begin{table}[h!]
\begin{center}
\begin{tabular}{| c | c | c | c | c | c | c | c | c | c | }
\hline
$\xi$ & $\frac{M}{m_{\rm Pl}}$ & $\phi_+/m_{\rm Pl}$ & $\phi_-/m_{\rm Pl}$ & \thead{$H_+$\\($10^{13}$ GeV)} & \thead{$H_-$\\($10^{13}$ GeV)} & \thead{$M_{I+}$\\($10^{13}$ GeV)} & \thead{$M_{I-}$\\($10^{13}$ GeV)} & $N_+$ & $N_-$ \\
\hline
\multirow{5}{*}{\rotatebox[origin=c]{90}{-0.001}}& 50 & 43.62 & 42.16 & 3.74 & 4.15 & 5.58 & 6.40 & 10.9 & 17.1 \\
& 70 & 63.57 & 62.07 & 3.67 & 4.04 & 5.27 & 5.93 & 10.9 & 17.1 \\
& 100 & 93.57 & 92.04 & 3.51 & 3.81 & 4.89 & 5.39 & 10.8 & 17.1 \\
& 200 & 193.71 & 192.16 & 3.03 & 3.21 & 4.08 & 4.35 & 10.8 & 17.0 \\
& 500 & 494.33 & 492.79 & 2.33 & 2.40 & 3.07 & 3.18 & 10.7 & 16.9 \\
\hline
\multirow{5}{*}{\rotatebox[origin=c]{90}{-0.002}}& 50 & 43.75 & 42.33 & 3.08 & 3.31 & 4.58 & 5.09 & 10.8 & 17.0 \\
& 70 & 63.74 & 62.29 & 2.95 & 3.14 & 4.23 & 4.60 & 10.8 & 17.0 \\
& 100 & 93.81 & 92.33 & 2.75 & 2.90 & 3.82 & 4.09 & 10.8 & 17.0 \\
& 200 & 194.18 & 192.68 & 2.30 & 2.38 & 3.09 & 3.22 & 10.7 & 16.9 \\
& 500 & 495.31 & 493.87 & 1.80 & 1.84 & 2.37 & 2.43 & 10.6 & 16.8 \\
\hline
\multirow{5}{*}{\rotatebox[origin=c]{90}{-0.003}}& 50 & 43.88 & 42.51 & 2.58 & 2.72 & 3.83 & 4.17 & 10.7 & 16.9 \\
& 70 & 63.93 & 62.51 & 2.45 & 2.57 & 3.49 & 3.74 & 10.7 & 16.9 \\
& 100 & 94.06 & 92.62 & 2.27 & 2.36 & 3.14 & 3.31 & 10.7 & 16.9 \\
& 200 & 194.62 & 193.17 & 1.91 & 1.96 & 2.55 & 2.64 & 10.7 & 16.8 \\
& 500 & 496.08 & 494.74 & 1.58 & 1.61 & 2.08 & 2.12 & 10.6 & 16.8 \\
\hline
\end{tabular}
\caption{Intermediate symmetry breaking at scale $M_I$ which generates monopoles with the yield ($Y_M=n_M/s$) corresponding to the MACRO bound $Y_M^+\simeq 10^{-27}$ (corresponding parameters are depicted with subscript `+'), and an adopted threshold for observability $Y_M^- = 10^{-35}$ (corresponding parameters are shown with subscript `-').  Inflation is driven by a real GUT singlet scalar with a Coleman-Weinberg potential and non-minimal coupling to gravity for different values of $\xi$ and $M$. We have shown the corresponding values of the inflaton field ($\phi$), Hubble parameter ($H$), and the number of $e$-foldings ($N$) experienced by the monopoles. We have set $T_r = 10^7$ GeV.}
\label{tab:mon-infl}
\end{center}
\end{table}
 Grand unified theories based on $SU(5)$ \cite{Georgi:1974sy}, $SO(10)$ \cite{Fritzsch:1974nn} or $E(6)$ \cite{Shafi:1978gg} predict the existence of topologically stable magnetic monopoles with masses about an order of magnitude larger than the unification scale \cite{tHooft:1974kcl,Polyakov:1974ek,Lazarides:1980cc,Shafi:1984wk}. The inflationary scenario we have discussed inflates away these GUT monopoles entirely. In $SO(10)$ and $E(6)$, topologically stable monopoles appear during the spontaneous breaking of the intermediate symmetry if the relevant homotopy group $\pi_2(\mathcal{M})$ of the vacuum manifold $\mathcal{M}$ is non-trivial. The MACRO experiment \cite{Ambrosio:2002qq} puts an upper bound  $2.8\times 10^{-16}$ ${\mathrm{cm}^{-2}\mathrm{s}^{-1}\mathrm{sr}^{-1}}$ on the monopole flux for monopole mass $m_M\sim 10^{14}$ GeV. This bound on the monopole flux implies that the upper bound on the comoving number density, the so-called yield, $Y_M^+ = n_M/s \simeq 10^{-27}$ \cite{Kolb:1990vq}, with $n_M$ being the monopole number density and $s$ the entropy density. We assume a lower bound for the threshold for observability $Y_M^{\rm -}\approx 10^{-35}$, corresponding to a monopole flux of $10^{-24}$ ${\mathrm{cm}^{-2}\mathrm{s}^{-1}\mathrm{sr}^{-1}}$. Table~\ref{tab:mon-infl} shows the number of $e$-folds corresponding to the monopole yields $Y_M^+$ and $Y_M^-$ along with the corresponding values of the inflaton field, Hubble parameter and the symmetry breaking scale for typical choices of the parameters $\xi$ and $M$ for successful inflation.

Topologically stable cosmic strings are produced if the fundamental homotopy group ($\pi_1(\mathcal{M})$) of the vacuum manifold $\mathcal{M}$ is non-trivial. The string tension corresponding to a spontaneous symmetry breaking achieved by a VEV $\left<\chi_{\rm str}\right>$ is expressed as \cite{Hill:1987qx,Hindmarsh:2011qj}
\begin{equation}\label{eq:Gmu}
G\mu\simeq\frac{1}{8}B(\frac{\lambda_{\mathrm{str}}}{g^2})\left(\frac{\left<\chi_{\rm str}\right>}{m_{\rm Pl}}\right)^2 ,
\end{equation} 
where $g$ is the effective gauge coupling constant and the function $B(x)$ is given by
\begin{align}\label{eq:Bx}
B(x)= \begin{cases} 
1.04 \ x^{0.195} & \mbox{for} \ 10^{-2}\lesssim x\lesssim 10^2 \\
2.4/\ln(2/x) & \mbox{for} \ x\lesssim 0.01.
\end{cases}
\end{align}
 
\begin{figure}[h!]
\begin{center}
\includegraphics[scale=0.7]{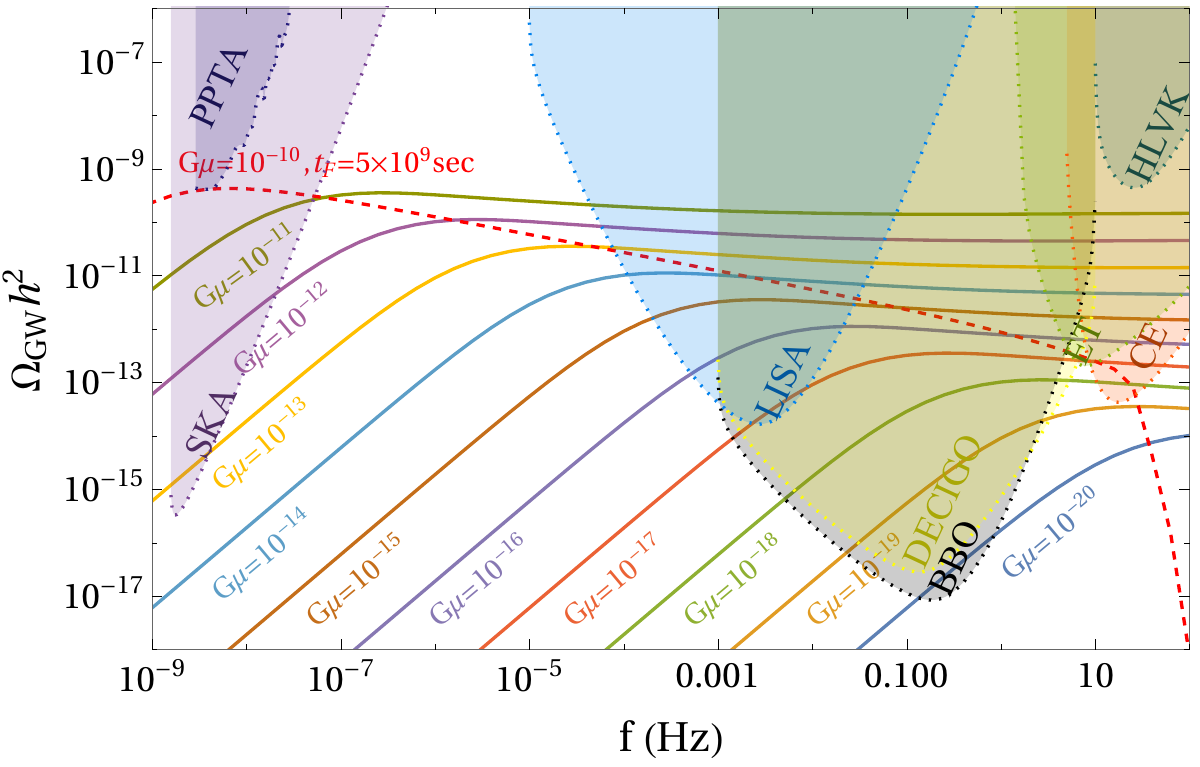}
\caption{Gravitational wave spectra from topologically stable cosmic strings for different values of the dimensionless string tension $G\mu$. We have shown the gravitational wave spectra for $G\mu$ varying from $10^{-20}-10^{-10}$. The gravitational wave background from strings with $G\mu \lesssim 4.6\times 10^{-11}$ that can form loops from a very early time ($t_F\sim 10^{-20}$ sec) satisfies the present PPTA exclusion limit \cite{Shannon:2015ect}. We have depicted the sensitivity curves \cite{Thrane:2013oya, Schmitz:2020syl} for proposed experiments, namely, SKA \cite{5136190, Janssen:2014dka}, CE \cite{PhysRevLett.118.151105}, ET \cite{Mentasti:2020yyd}, LISA \cite{Bartolo:2016ami, amaroseoane2017laser}, DECIGO \cite{Sato_2017}, BBO \cite{Crowder:2005nr, Corbin:2005ny}, HLVK \cite{KAGRA:2013rdx}. The solid lines represent the stochastic gravitation wave backgrounds from topologically stable strings with $G\mu$ ranging from $10^{-20}$ to $10^{-11}$ and horizon reentry time $t_F\sim 10^{-20}$ sec. The red dashed line shows the gravitational wave background from cosmic strings with $G\mu = 10^{-10}$ and horizon reentry time $t_F \simeq 5\times 10^9$ sec.}\label{fig:GWs}
\end{center}
\end{figure} 
\begin{figure}[h!]
\begin{center}
\includegraphics[scale=0.75]{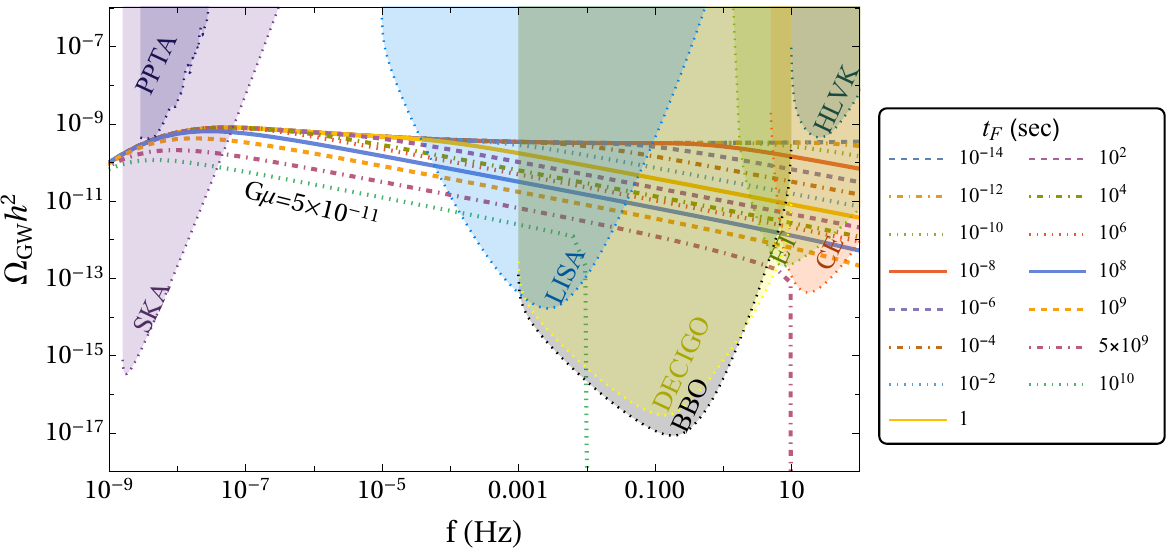}
\caption{Gravitational wave spectra from topologically stable cosmic strings with typical dimensionless string tension $G\mu=5\times 10^{-11}$, for different values of the horizon reentry time $t_F$.}\label{fig:GWs_varying_tF}
\end{center}
\end{figure} 
\begin{figure}[h!]
\centering
\includegraphics[scale=0.7]{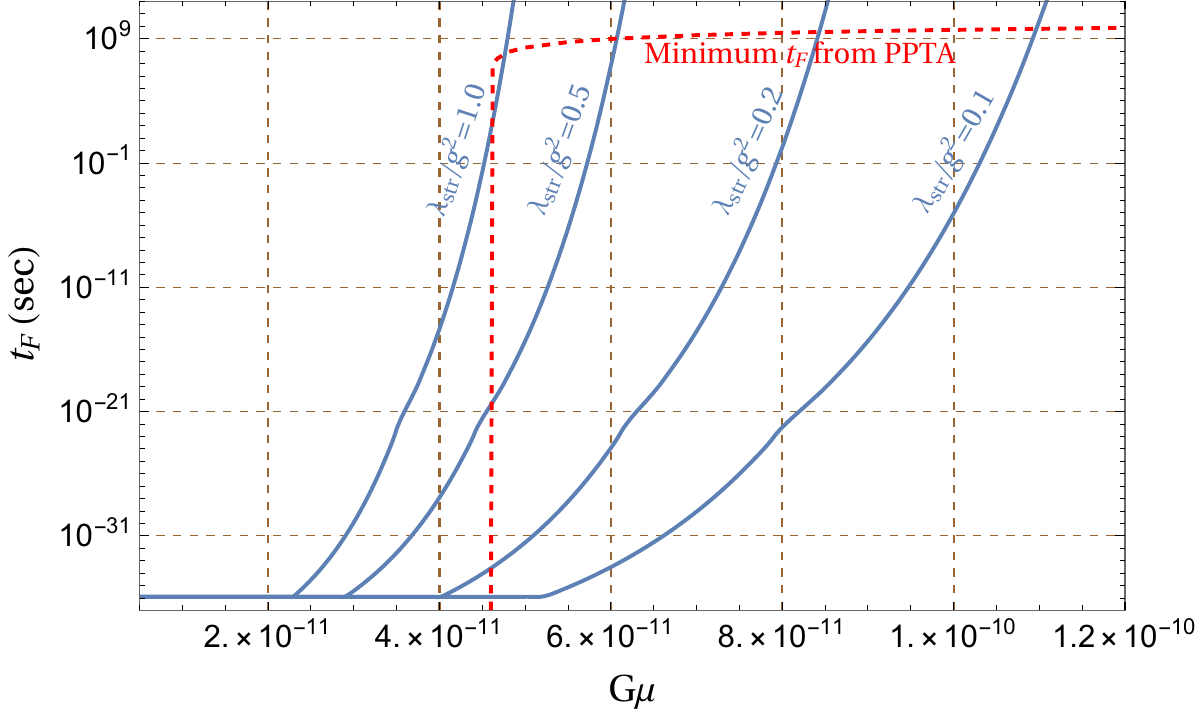}
\caption{Horizon reentry time $t_F$ of cosmic strings as a function of $G\mu$ for different choices of $\lambda_{\rm str}/g^2$ (blue curves). The red dashed line shows the minimum $t_F$ values that satisfy the PPTA bound.}
\label{fig:tF-gmu}
\end{figure}
The mean inter-string separation at the time of formation is $p~\xi_G$, where $p\simeq 2$ is a geometric factor \cite{Lazarides:2019xai,Chakrabortty:2020otp}. This distance increases by a factor $\exp(N_{\rm str})$ during inflation, where $N_{\rm str}$ is the number of $e$-foldings experienced by the strings. The latter can reenter the post-inflationary horizon at time $t_F$ when the mean inter-string separation becomes equal to the particle horizon $d_{\rm hor}=2t_F \ (3t_F)$ in the radiation (matter) dominated universe. More specifically, the strings reenter the horizon at a time $t_F$ after reheating for
\begin{equation}\label{inter-string-dist}
d_{\rm hor}=p~\xi(\phi_{\mathrm{str}})\exp(N_{\mathrm{str}})\left(\frac{t_r}{\tau}\right)^{\frac{2}{3}} 
\frac{T_r}{T}, 
\end{equation}
where the temperature $T$ at $t_F$ during the radiation dominated universe is given by
\begin{equation}\label{time-rad-dom}
T^2=\sqrt{\frac{45}{2\pi^2g_*}}\,\frac{m_{\rm Pl}}{t_F} .
\end{equation}

The strings intercommute and form loops soon after their horizon reentry, which subsequently decay via gravitational wave radiation. Fig.~\ref{fig:GWs} shows the stochastic gravitational wave background radiated by the string loops with $G\mu$ varying between $10^{-20}-10^{-10}$.  We have depicted the sensitivity curves \cite{Thrane:2013oya, Schmitz:2020syl} for some of the proposed experiments, namely, SKA \cite{5136190, Janssen:2014dka}, CE \cite{PhysRevLett.118.151105}, ET \cite{Mentasti:2020yyd}, LISA \cite{Bartolo:2016ami, amaroseoane2017laser}, DECIGO \cite{Sato_2017}, BBO \cite{Crowder:2005nr, Corbin:2005ny}, HLVK \cite{KAGRA:2013rdx}, which could detect this gravitational wave background. We have computed the gravitational wave spectra following Ref.~\cite{Lazarides:2022spe} and Refs.~\cite{Olmez:2010bi,Cui:2019kkd,Auclair:2019wcv,Blanco-Pillado:2013qja,Blanco-Pillado:2017oxo} therein. Gravitational waves from cosmic strings and from various hybrid defects have also been studied Refs.~\cite{Vachaspati:1984gt,Martin:1996ea,Martin:1996cp,Vilenkin:2000jqa,Leblond:2009fq,Sousa:2013aaa,Blanco-Pillado:2017oxo,Cui:2017ufi,Cui:2018rwi,Guedes:2018afo,Gouttenoire:2019kij,Buchmuller:2019gfy,King:2020hyd,Ellis:2020ena,Buchmuller:2020lbh,King:2021gmj,Buchmuller:2021dtt,Buchmuller:2021mbb,Masoud:2021prr,Dunsky:2021tih,Chun:2021brv,Afzal:2022vjx,Ahmed:2022rwy,Lazarides:2022jgr}. The gravitational wave background from strings with $G\mu \lesssim 4.6\times 10^{-11}$ that can form loops from a very early time ($t_F\sim 10^{-20}$) satisfies the present PPTA exclusion limit \cite{Shannon:2015ect,Lazarides:2021uxv}. The stochastic gravitation wave backgrounds from topologically stable strings with $G\mu$ ranging from $10^{-20}$ to $10^{-11}$ are computed assuming a horizon reentry time $t_F\sim 10^{-20}$ sec without loss of generality.  The red dashed line in Fig.~\ref{fig:GWs} shows the gravitational wave background from cosmic strings with $G\mu = 10^{-10}$ and $t_F \simeq 5\times 10^9$ sec which is consistent with the PPTA bound. This is because strings reentering the particle horizon at time $t_F$ mostly populate loops of size $\alpha t_i$ ($\alpha \sim 0.1$ and $t_i\gtrsim t_F$), and the formation of smaller loops will be less probable \cite{Blanco-Pillado:2013qja,Blanco-Pillado:2017oxo}. Therefore, the contribution to the gravitational wave background in the higher frequency region will be reduced. Fig.~\ref{fig:GWs_varying_tF} shows the variation of the gravitational wave background for different choices of $t_F$ starting from $10^{-14}$ sec to $10^{10}$ sec and a typical value of $G\mu = 5\times 10^{-11}$. Needless to say, the sharp fall at the highest frequency in the spectrum for a specific $t_F$ is because the gravitational wave bursts beyond this frequency are too infrequent to form an unresolved gravitational wave background, namely, the burst-rate becomes lower than the frequency \cite{Cui:2019kkd}. Fig.~\ref{fig:tF-gmu} depicts the minimum $t_F$ required for the gravitational wave spectra to satisfy the PPTA bound and the horizon reentry time ($t_F$) of the cosmic strings in our inflationary model for typical choices of $\lambda_{\rm str}/g^2 = 1, 0.5, 0.2$ and $0.1$ as a function of $G\mu$. 

\section{Topologically Stable Monopoles and Cosmic Strings in $SO(10)$}
\label{sec:so10}
\begin{figure}[h!]
\centering
\includegraphics[width=0.9\textwidth]{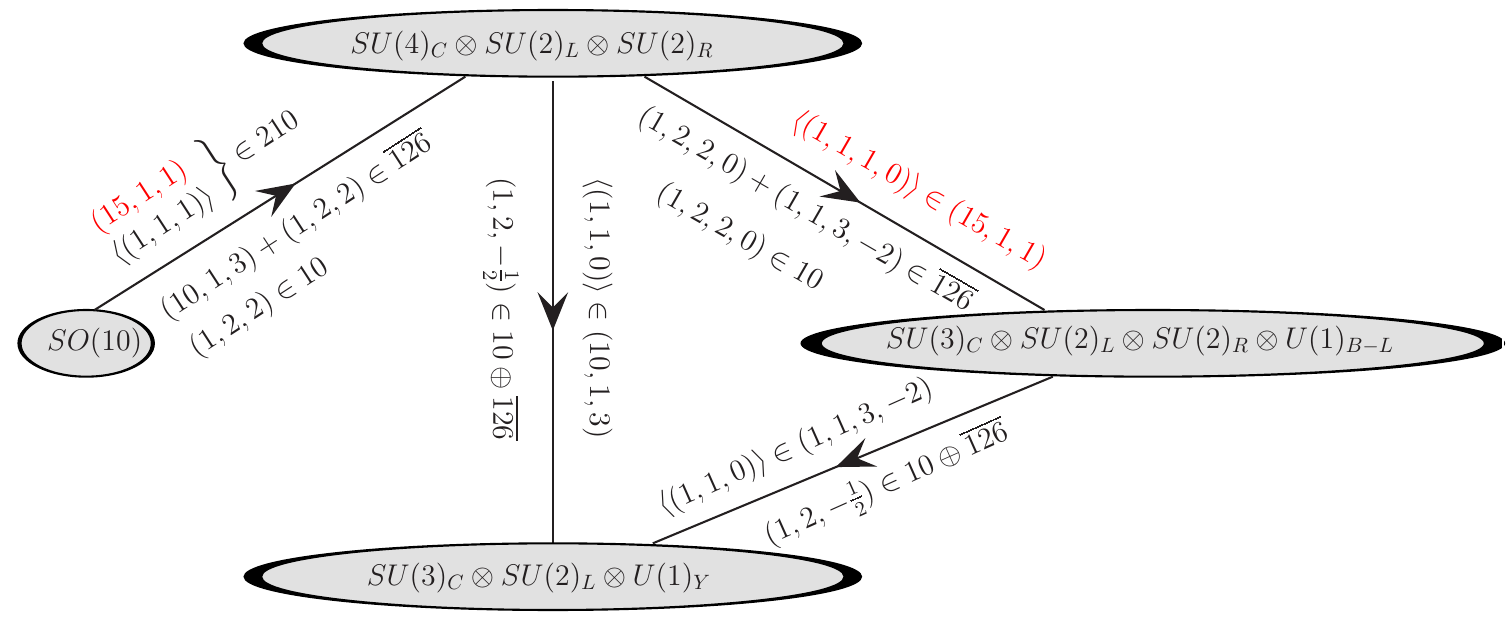}
\caption{Breaking patterns of $SO(10)$ and the relevant scalar multiplets.}
\label{fig:breaking-patterns}
\end{figure}
In this section we discuss two realistic symmetry breaking patterns of $SO(10)$ which predict the formation of topologically stable monopoles and cosmic strings at the intermediate scale. The VEV of $210$ breaks $SO(10)$ to the gauge symmetry, $SU(4)_C\times SU(2)_L\times SU(2)_R$ \cite{Pati:1974yy}. This breaking produces topologically stable GUT scale monopoles \cite{Lazarides:1980cc,Lazarides:2019xai} which are inflated away in our inflationary scenario. A non-zero VEV for $(10,1,3)$ in $\overline{126}$ spontaneously breaks $SU(4)_C\times SU(2)_L\times SU(2)_R$ group to the SM. In this case, topologically stable monopoles and cosmic strings are formed at the same intermediate scale. In the second case, a VEV for $(15,1,1)\in 210$ breaks $SU(4)_C\times SU(2)_L\times SU(2)_R$ to $SU(3)_C\times SU(2)_L\times SU(2)_R\times U(1)_{B-L}$, generates intermediate scale monopoles with mass related to $M_I$. The latter group is then broken to the SM via the VEV of $(1,1,3,-2)\in (10,1,3)$ of $\overline{126}$, and topologically stable strings with mass scale related to $M_{II}$ are formed during this phase transition. The SM gauge symmetry is broken by the VEV of the Higgs doublet which comes from a combination of the bi-doublets in $\overline{126}$- and $10$-dimensional scalars.

\begin{figure}[htbp]
\begin{center}
\includegraphics[scale=0.4]{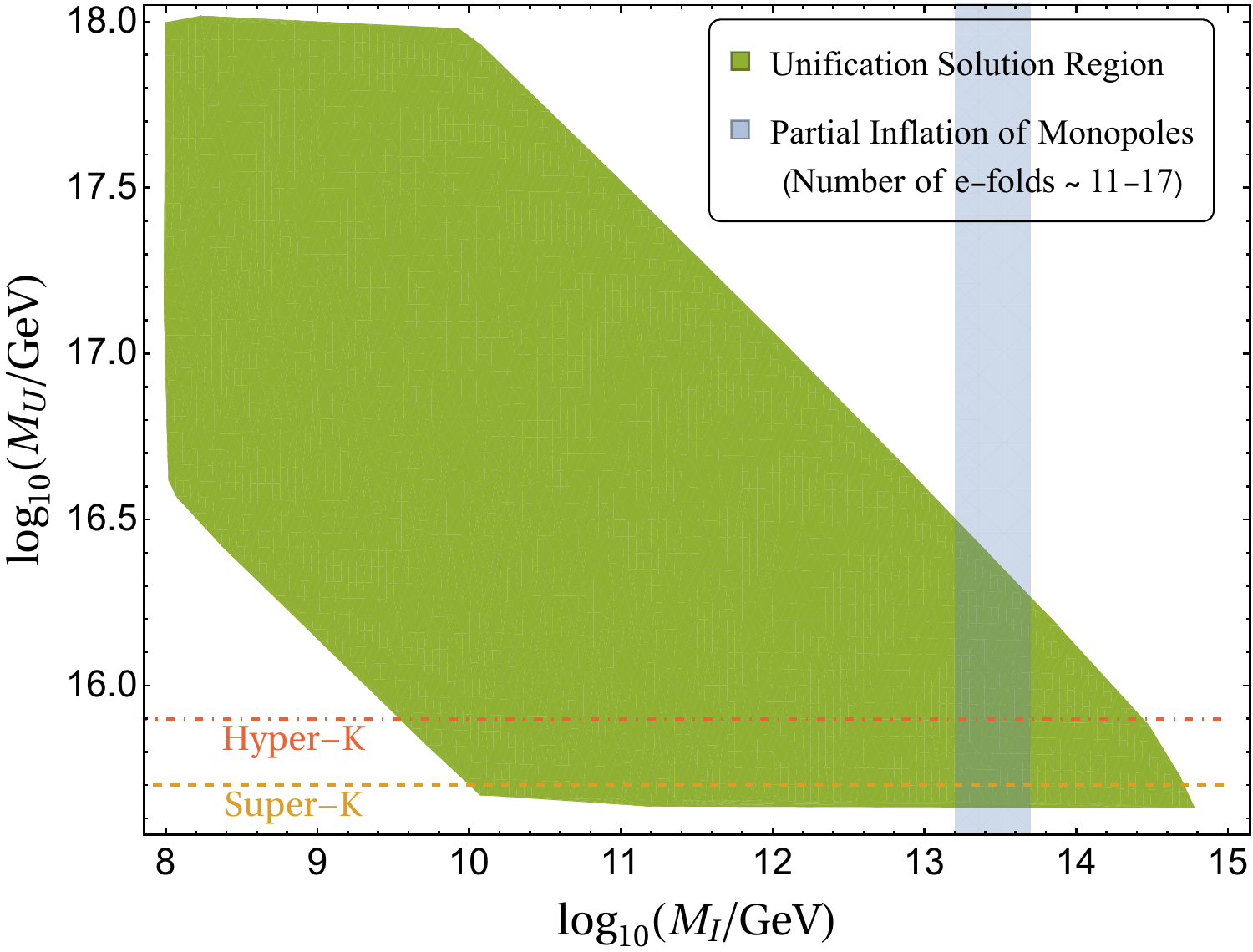}
\caption{Unification solutions for the breaking chain $SO(10)\xrightarrow{M_U} SU(4)_C\times SU(2)_L\times SU(2)_R\xrightarrow{M_I} SU(3)_C\times SU(2)_L\times U(1)_Y$. We have set $R\in [1/2,2]$. The horizontal dashed and dot-dashed lines depict the Super-Kamiokande exclusion limit and Hyper-Kamiokande sensitivity on the unification scale $M_U$. The vertical shaded region shows the intermediate scale $M_I$ associated with the monopoles which undergo around $11$ to $17$ $e$-foldings. In this range the monopole flux satisfies the MACRO bound and is within the adopted threshold for observation (see Table~\ref{tab:mon-infl}).}
\label{fig:g224}
\end{center}
\end{figure} 
\begin{figure}[htbp]
\begin{center}
 \subfloat[][Contour plot of $M_U$ as a function of $M_I$ and $M_{II}$.]{\includegraphics[scale=0.65]{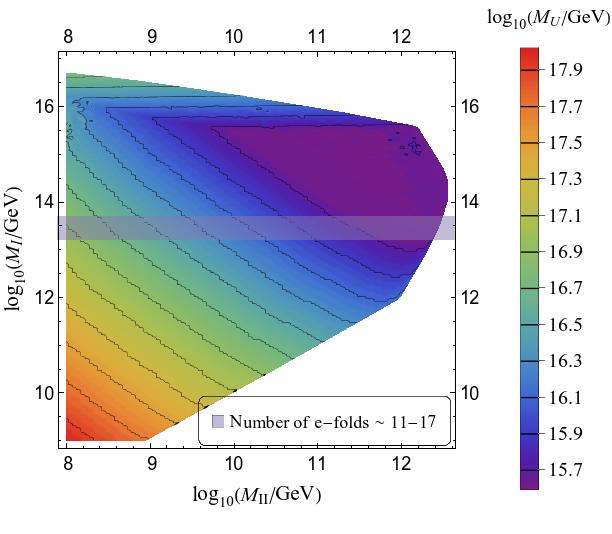}\label{fig:g224-g2231-a}} \\
 \subfloat[][$M_U$ versus $M_I$ plot.]{\includegraphics[width=0.48\textwidth]{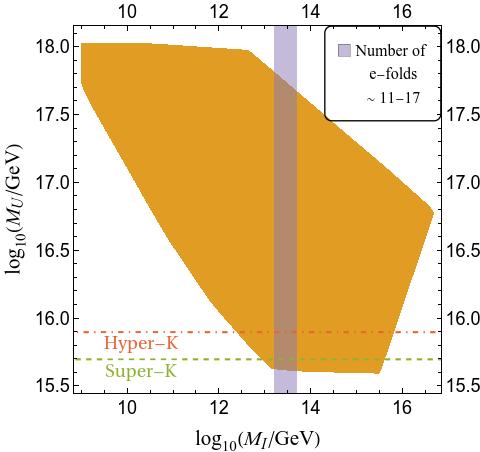}\label{fig:g224-g2231-b}}
 \subfloat[][$M_U$ versus $M_{II}$ plot.]{\includegraphics[width=0.48\textwidth]{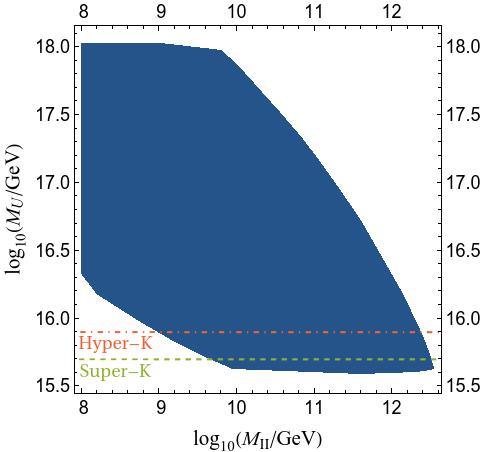}\label{fig:g224-g2231-c}}
\caption{Unification solutions for the breaking chain $SO(10)\xrightarrow{M_U} SU(4)_C\times SU(2)_L\times SU(2)_R \xrightarrow{M_I} SU(3)_C\times SU(2)_L\times SU(2)_R\times U(1)_{B-L}\xrightarrow{M_{II}} SU(3)_C\times SU(2)_L\times U(1)_Y$. $R\in [1/2,2]$ in this case. The horizontal dashed and dot-dashed lines in (b) and (c) depict the Super-Kamiokande exclusion limit and Hyper-Kamiokande sensitivity on the unification scale $M_U$. The horizontal and vertical shaded regions in (a) and (b) respectively, show the intermediate scale $M_I$ associated with the monopoles which undergo around $11$ to $17$ $e$-foldings. In this range the monopole flux satisfies the MACRO bound and is within the adopted threshold for observation (see Table~\ref{tab:mon-infl}).}
\label{fig:g224-g2231}
\end{center}
\end{figure}

The light scalar multiplets which contribute to the renormalization group evolution of the gauge couplings are the ones acquiring VEVs during the associated and subsequent gauge symmetry breakings in accordance with the \textit{extended survival hypothesis}. Fig.~\ref{fig:breaking-patterns} shows two breaking patterns along with the scalar multiplets. To obtain the unification solutions we assume that the heavy scalars in a parent multiplet have masses within the ratio $R$ with respect to the masses of heavy gauge bosons \cite{Babu:2015bna, Chakrabortty:2019fov,Meloni:2019jcf}. We have followed Ref.~\cite{Jones:1981we} to compute the two loop $\beta$-coefficients, Refs.~\cite{WEINBERG198051,Hall:1980kf} to obtain the matching conditions, and Refs.~\cite{Chakrabortty:2019fov,Chakrabortty:2017mgi} and the references therein to obtain the unification solutions compatible with the electroweak observables.

 Fig.~\ref{fig:g224} shows the unification  scale $M_U$ as function of the intermediate scale $M_I$ for $R\in [1/5,5]$ for the single intermediate scale breaking pattern. We have taken $M_U$ upto $10^{18}$ GeV and $M_I$ starting from $10^8$ GeV. The solution regions are obtained within  $\log_{10}(M_I/\mathrm{GeV})\in [8.0, 14.8]$, $\log_{10}(M_U/\mathrm{GeV})\in [15.6, 18.0]$ and the unified coupling $g_U\in [0.56, 0.71]$. The constraint $\tau(p\to e^+\pi^0)\geq 2.4\times 10^{34}$ from the Super-Kamiokande (Super-K) experiment \cite{Super-Kamiokande:2020wjk} implies that the unification scale $\log_{10}(M_U/{\rm GeV}) \gtrsim 15.7$. The Hyper-Kamiokade (Hyper-K) experiment, on the other hand, will probe $\tau(p\to e^+\pi^0)$ upto $10^{35}$ yrs within $3\sigma$ confidence level \cite{Dealtry:2019ldr}, which corresponds to the unification scale $\log_{10}(M_U/{\rm GeV}) \simeq 15.9$. We have shown the Super-K exclusion limit and the Hyper-K sensitivity in Figs.~\ref{fig:g224} and \ref{fig:g224-g2231}.

For the case with two intermediate symmetry breaking scales, we have set $R\in [1/2,2]$. Fig.~\ref{fig:g224-g2231} shows a plot of $M_U$ as a function of the intermediate scales $M_I$ and $M_{II}$, and separate plots of $M_U$ versus $M_I$ and $M_U$ versus $M_{II}$. We have again varied $M_U$ upto $10^{18}$ GeV, with $M_I$ starting from $10^9$ GeV and $M_{II}$ from $10^8$ GeV. The solution regions are within $\log_{10}(M_{II}/\mathrm{GeV})\in [8.0, 12.6]$,  $\log_{10}(M_I/\mathrm{GeV})\in [9.0, 16.7]$, $\log_{10}(M_U/\mathrm{GeV})\in [15.6, 18.0]$ and the unified coupling $g_U\in [0.522,0.694]$. The primordial monopole flux is consistent with the MACRO bound if the corresponding symmetry breaking during inflation undergoes at least $11$ $e$-foldings, and it could be within the adopted threshold for observability if the $e$-folding number is less than $17$ (see Table~\ref{tab:mon-infl}). The relevant intermediate scale lies within $13.2\lesssim\log_{10}(M_I/\mathrm{GeV})\lesssim 13.7$ as can be seen in Table~\ref{tab:mon-infl}. We have highlighted this region in Figs.~\ref{fig:g224} and \ref{fig:g224-g2231}. It is exciting that there are solution regions that will be probed by Hyper-K and also provide partially inflated monopoles within the adopted threshold for observability.
\section{Conclusions}
\label{sec:summary}
We have explored the implications for topologically stable magnetic monopoles and cosmic strings associated with symmetry breaking phase transitions that encountered a limited number of $e$-foldings during inflation. The latter is driven by a Coleman-Weinberg potential and the GUT singlet scalar inflaton field has non-minimal coupling to gravity. The predicted scalar spectral index $n_s$ lies within the one sigma region determined by  Planck 2018 + BICEP2/Keck Array (BK18) experiments + BAO, and the prediction $ r \gtrsim 0.003$ for the tensor to scalar ratio is  within the reach of ongoing and planned experiments \cite{SimonsObservatory:2018koc,Abazajian:2019eic}. We provide two realistic examples based on $SO(10)$ that yield both monopoles and cosmic strings in the observable range. The impact of partial inflation on the gravitational wave spectrum emitted by strings is discussed. The observations of monopoles and gravitational waves from strings would shed light on the symmetry breakings that took place in the early universe.

Our discussion can be extended to composite topological structures such as monopoles connected to strings and walls bounded by strings that experience a limited number of $e$-foldings during inflation. A good example is provided by the domain walls bounded by strings in axion models in which the axion strings experience some inflation before combining with the domain walls. Such extended structures may provide additional seeds for galaxy formation \cite{Lazarides:1984pq}.
\section{Acknowledgment}
We thank Vedat Nefer Şenoğuz for a careful reading of our manuscript and helpful suggestions for improvement.

\bibliographystyle{mystyle}
\bibliography{GUT_TD}

\end{document}